\documentclass[twocolumn,showpacs,pre,floatfix,amsmath,amssymb,notitlepage]{revtex4-1}
\usepackage{epsfig}
\usepackage{subfigure}
\usepackage{amsmath}
\usepackage{color}
\usepackage{amssymb}
\usepackage{setspace}
\usepackage{graphicx}
\usepackage{dcolumn}
\usepackage{bm}
\usepackage{times}
\usepackage{enumerate}
\usepackage{footnote}
\usepackage{upgreek}
\input epsf

\graphicspath{{figures/}}

\begin{document}
	
	\author{Linpeng Gu$^{1}$}
	\author{Binbin Wang$^{1}$}
	\author{Qingchen Yuan$^{1}$}
	\author{Liang Fang$^{1}$}
	\author{Qiang Zhao$^{2}$}
	\author{Xuetao Gan$^{1}$, Jianlin Zhao$^{1}$}
    \email{zhaoqiang@qxslab.cn; xuetaogan@nwpu.edu.cn} 
	\affiliation{$^{1}$Key Laboratory of light-field manipulation and information acquisition, Ministry of Industry and Information Technology, and Shaanxi Key Laboratory of Optical Information Technology, School of Science, Northwestern Polytechnical University, Xi'an 710072, China}
	\affiliation{$^{2}$Qian Xuesen laboratory of Space Technology, China Academy of Space Technology, Beijing 100094, China}
	\date{\today}
	
	\title{Fano resonance from a one-dimensional topological photonic crystal}

\maketitle

{\bf
An ultra-compact one-dimensional topological photonic crystal (1D-TPC) is designed in a single mode silicon bus-waveguide to generate Fano resonance lineshape. The Fano resonance comes from the interference between the discrete topological boundary state of the 1D-TPC and the continuum high-order leaky mode of the bus-waveguide. Standalone asymmetric Fano resonance lineshapes are obtained experimentally in the waveguide transmission spectrum with a maximum extinction ratio of 33 dB and a slope ratio of 10 dB/nm over a broadband flat background.  }\\

\section{Introduction}\label{sec:01}
Constructive and destructive interferences between localized discrete state and continuum state usually give rise to asymmetric spectra, which is known as Fano resonance~\cite{fano1961effects}. Asymmetric lineshape of Fano resonance exhibits an ultra-sharp variation from the minimum to the maximum compared to the symmetric Lorenzian lineshape with an approximative quality ($Q$) factor~\cite{yu2016all}. This steeper resonant peak provides great advantages for chip-scale integrated applications like low-threshold lasing~\cite{limonov2017fano,yu2017demonstration}, high-sensitivity optical sensing~\cite{yi2010highly,tu2017high,zhang2020photonic} and low-power all-optical switching~\cite{yu2016all,nozaki2013ultralow,yu2014fano,heuck2013improved}. Thanks to the maturity and rapid development of planar processing technology, a large number of structures such as microring resonators~\cite{zhang2016optically,qiu2012asymmetric,gu2019compact,gu2020fano,fang2020controlling}, microdisks~\cite{li2012experimental} and photonic crystal (PC) nano-cavities\cite{yu2016all,yu2017demonstration,nozaki2013ultralow,yu2014fano,heuck2013improved} have been proposed to generate the Fano resonance. Among them, more compact designs are always expected to meet the rapid development of large-scale photonic integrated circuits.

Recently, topological photonics attracted remarkable attention due to its unique properties and made great progress especially when it is combined with the mature CMOS-compatible platforms. Photonic analogues of topological quantum systems can be easily realized via PC or microring arrays~\cite{raghu2008analogs,chen2017valley,khanikaev2017two}, which promise a new generation of compact chip-scale photonic devices. For example, topological boundary states (TBSs) existing in open bandgaps of PCs become the focus in building high $Q$ factor nano-cavities~\cite{ota2018topological,zhang2020low,shao2020high,han2019lasing}. A topology-based design of zero-dimensional interface state in a one-dimensional (1D) PC nanobeam guarantees the existence of only one cavity mode within its photonic bandgap~\cite{ota2018topological}. This special design takes advantage of highly localized mode in an ultra-small volume with a high energy density in the nano-cavity. In addition, a dislocation-based topological cavity mode is convinced to be more stable against perturbations compare with conventional photonic crystal defect cavities\cite{li2018topological}. Besides, depending on the characteristics of the topology of bulk bands, topologically protected edge state can transport against certain classes of defects~\cite{shalaev2019robust,he2019silicon}. By replacing the localized discrete state in Fano resonance with a TBS, topological protection in photonics is promising to offer new prospects for guiding and manipulating Fano resonance lineshapes as well.

For example, by coupling a Fabry-Perot cavity mode with a topological edge mode, it is possible to construct a high $Q$ Fano resonance in a compact one-dimensional topological photonic crystal (1D-TPC) heterostructure~\cite{gao2018fano}. Fano resonance that aroused by a topological high-$Q$ dark edge mode and a low-$Q$ bright edge mode is different from the trivial cases, which are commonly sensitive to geometry~\cite{zangeneh2019topological}. The ultra-sharp spectra of topologically protected Fano resonances can be guaranteed by designs without stringent geometrical requirements, and with a complete immunity to structural disorders~\cite{zangeneh2019topological,wang2020robust}. By coupling the valley-dependent topological edge states with a double-degenerate cavity~\cite{ji2020topologically}, the immune property to system impurities has also been verified on the two-dimentional photonic valley Hall insulators. These efforts pave a way for the topologically protected robust Fano devices with ultra-compact size such as optical switches, low-threshold nanolasers, and ultra-sensitive optical sensors.

In this letter, we demonstrate the generation of Fano resonance through an ultra-compact 1D-TPC structure embedded in a single mode bus-waveguide. Finite difference time domain (FDTD) method and transfer matrix method are utilized to analyze the designed structure and the Fano resonance lineshapes. The localized discrete state and continuum state of the constructed Fano resonance are the highly localized TBS mode of 1D-TPC and high-order leaky mode of the bus-waveguide, respectively. The results are further verified by experimentally fabricating the devices in a silicon slab. 

\begin{figure*}
	\centering
	\includegraphics[width=\linewidth]{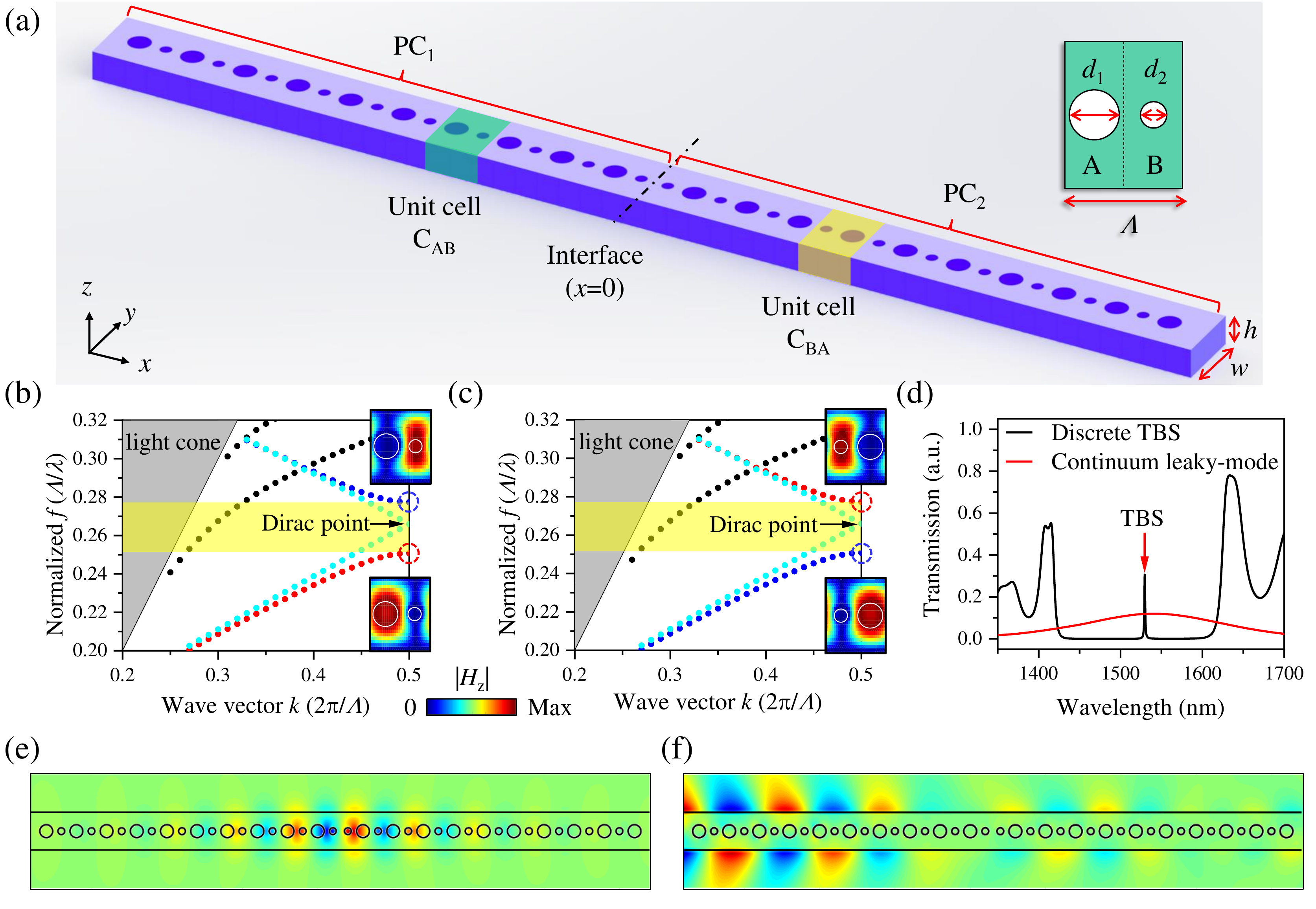}
	\caption{ (a) Schematic image of the designed 1D-TPC embedded in a silicon bus-waveguide. Inset is the top view of the unit cell C$_{\text{AB}}$. (b) and (c) Band diagrams of PC$_{1}$ and PC$_{2}$ with $w=500$ nm, $h=220$ nm, $d_1=170$ nm, $d_2=90$ nm and $\Lambda=400$ nm, respectively. Band inversion occurs at lowest two bands when flipping the unit cells. Inset shows the intensity distributions of the band edge states under the light cone (the grey area). (d) Simulated transmission spectra of the discrete TBS of the 1D-TPC and continuum high-order leaky mode of the bus-waveguide in the structure shown in (a). (e) and (f) Calculated electrical field $E_y$ profiles of the resonant transmission of TBS and the high-order leaky mode passing through the 1D-TPC region. }
	\label{fig:fig1}
\end{figure*}

\section{Design of 1D-TPC}
The proposed 1D-TPC embedded in a silicon bus-waveguide is first designed and simulated by using FDTD technique. The schematic image is shown in Fig.~\ref{fig:fig1}(a) with a waveguide width $w$ and height $h$. The structure with periodic air-holes in a bus-waveguide is similar as that in traditional PC nanobeam cavity, which forms resonance modes in the defect of a 1D PC with the same air-hole diameter. Differently, the 1D-TPC is composed of two halves of 1D-PCs, named by PC$_{1}$ and PC$_2$. PC$_{1}$ includes periodically stacked unit cells (C$_{\text{AB}}$, highlighted in dark green), which contains two inequivalent circular air-holes. The air-holes are with diameters of $d_1$ and $d_2$ ($d_1>d_2$), respectively, as shown in the upper-right inset of Fig.~\ref{fig:fig1}(a). PC$_{2}$ is mirror symmetric with PC$_{1}$, which comprises the inverted unit cells C$_{\text{BA}}$ (in yellow). C$_{\text{AB}}$ and C$_{\text{BA}}$ have the same period length $\Lambda$.

With a specific spatial inversion symmetry along the transmission direction of the bus-waveguide embedded with the 1D-TPC, a topological phase transition arising from band crossing can be realized as an analogy to the classic Su-Schrieffer-Heeger (SSH) model~\cite{su1979solitons}. This is reported to relate to the distinct geometric (Zak) phase of the PC structures on both sides~\cite{xiao2014surface,ota2018topological}. A guaranteed existence of the localized interface state can therefore be obtained by constructing an interface (at $x=0$) with PC$_1$ on the left and PC$_2$ on the right. Then, an ultra-compact 1D-TPC with the strongly localized TBS could be realized.  

To confirm this, by setting $w=500$ nm, $h=220$ nm, $d_1=170$ nm, $d_2=90$ nm and $\Lambda=400$ nm, we obtain the calculated band structures of PC$_{1}$ and PC$_{2}$ in the telecom-band range, respectively. Detailed images are plotted in Figs.~\ref{fig:fig1}(b) and (c). The frequency and wave vector in the band structure are plotted in dimensionless units as commonly used for scalable results\cite{joannopoulos2008molding}. The lowest two bands are marked with red and blue, respectively, as they have different geometric phases~\cite{xiao2014surface,ota2018topological}. The cyan bands, however, refer to the PC with air-hole diameters of such as $d_1=d_2=130$ nm. For the cells with different air-hole sizes, PC$_{1}$ and PC$_{2}$ share a similar band structure as a pair of inverted units. A wide photonic bandgap appears in the telecom-band between the lowest ($f_\text{norm}$=0.251 $\Lambda/\lambda$) and the second lowest ($f_\text{norm}$=0.277 $\Lambda/\lambda$) bands, which originates from Bragg interference of specific periodic refractive index distributions. 

However, if we gradually reduce $d_1$ while increasing $d_2$, then the photonic bandgap will gradually narrow. For the critical point when $d_1=d_2$, Dirac point appears at a frequency of 0.266 $\Lambda/\lambda$, as implied by the arrows in Figs.~\ref{fig:fig1}(b) and (c). That is, the band goes through the process of closing, crossing, and reopening again until $d_2>d_1$, which indicates a topological phase transition during the whole process~\cite{xiao2014surface}. The exchanged mode profile of the lowest two bands at the edges are given in the inset of Figs.~\ref{fig:fig1}(b) and (c) as well to confirm the existence of band inversion.

By further constructing PC$_1$ and PC$_2$ together, the TBS mode is observed from its transmission spectrum, as plotted in Fig.~\ref{fig:fig1}(d). Here, PC$_1$ and PC$_2$ are both designed with finite length of 10 unit cells. By scanning over the entire stopband, only one resonance peak is obtained as expected at the wavelength of 1529.6 nm with a $Q$ factor of 1,500. To be noted here, by increasing the number of unit cells and using a finer simulation grid, a higher $Q$ factor can be further obtained. This mode is located close to the center of the stopband and is far away from the band edge. It would be helpful to provide a new way with more freedom to avoid the control of mode number in a traditional multi-mode PC nanobeam cavity~\cite{xie2021efficient}. The electrical field $E_y$ profile of this standalone resonant mode is presented in Fig.~\ref{fig:fig1}(e). This TBS mode is strongly localized with a small volume of about 0.53$(\lambda/n)^3$. 

To realize the topology-based Fano resonance, an extra quasi-continuum or continuum state is needed to interact with this discrete TBS~\cite{fano1961effects}. Fortunately, a high-order quasi-transverse electric leaky mode of the silicon bus-waveguide can be excited through a laterally tapered waveguide structure~\cite{mehta2013fano}, which has a transmission spectrum shown in Fig.~\ref{fig:fig1}(d). Detailed electrical field $E_y$ profile of this high-order mode is presented in Fig.~\ref{fig:fig1}(f). The role of this leaky wave is to function as the continuum state. Thus, the resonant TBS of 1D-TPC can interfere with the excited high-order leaky mode and finally results in the asymmetric Fano lineshapes in the transmission spectrum. The whole process will be then analyzed by transfer matrix method and confirmed in our experiments.

\section{transfer matrix method}
\begin{figure}
	\centering
	\includegraphics[width=\linewidth]{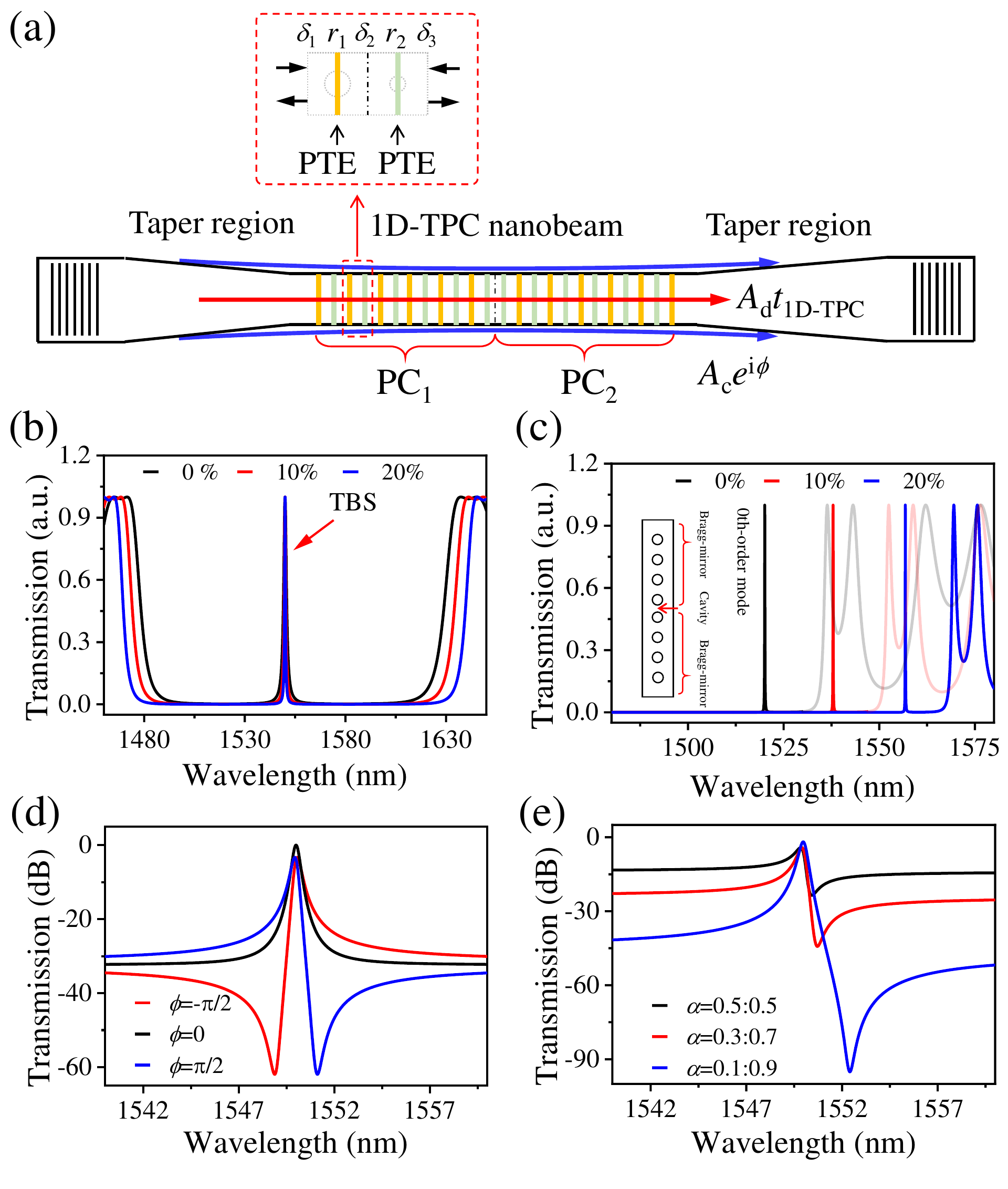}
	\caption{ (a) Transfer matrix model of the proposed 1D-TPC with laterally tapered waveguide structure. Inset shows transfer matrix model of a unit cell. (b) Calculated transmission spectra of TBS in the 1D-TPCs with increased reflectivity of the air-holes. (c) Calculated transmission spectra of the resonance mode in a traditional Bragg-reflection based PC nanobeam cavity with increased reflectivity of the air-holes. Band edge of the black and red lines are faded for clear vision. Inset gives the schematic image of the traditional PC nanobeam cavity. (d) Calculated Fano lineshapes with variable phase $\phi$ between the discrete TBS and continuum leak-mode state. (e) Calculated Fano lineshapes with variable amplitude ratio $\alpha$ between the discrete and continuum state.}
	\label{fig:fig2}
\end{figure}

To facilitate the analysis using the transfer matrix method, we schematically display the transfer matrix model of the proposed 1D-TPC structure in Fig.~\ref{fig:fig2}(a). PC$_1$ and PC$_2$ discussed above constitute the 1D-TPC within the bus-waveguide. Laterally tapered waveguides are directly connected to the 1D-TPC on both sides. High-order mode (blue arrows) of the bus-waveguide is transmitted as a leaky wave through the 1D-TPC region to interfere with the inside discrete TBS mode (red arrow).

For the proposed 1D-TPC, the air-holes in a unit cell could be viewed as two partially reflecting elements (PTEs) with distinct amplitude reflection coefficient of $r_1$ and $r_2$,  as displayed in the inset of Fig.~\ref{fig:fig2}(a). Thus, the transmission coefficient of the 1D-TPC could be easily calculated by transfer matrix method. For example, the transfer matrix $M_{\text{C}_\text{AB}}$ of unit cell C$_{\text{AB}}$ can be described as 
\begin{eqnarray}
\begin{array}{c}
M_{\text{C}_\text{AB}}=M_{w_1}M_{h_1}M_{w_2}M_{h_2}M_{w_3},
\\
\\
M_{h_p}=\frac{1}{i\sqrt{1-{r_p}^2}}\begin{bmatrix}
{-1}&{-r_p}\\
{r_p}&1
\end{bmatrix}, (p=1,2),
\\
\\
M_{w_q}=\begin{bmatrix}
{e^{i\delta_q}}&{0}\\
{0}&{e^{-i\delta_q}}
\end{bmatrix}, (q=1,2,3),
\end{array}
\end{eqnarray}
where $M_{h_p}$ is the transfer matrix of the PTE for the larger and small air-holes. $M_{w_q}$ respectively represents the transfer matrix of the three waveguide-segments beside the two air-holes, which has phase shift $\delta_q(q=1,2,3)$ of the guiding wave. Similarly, the transfer matrix of the inversion unit C$_{\text{BA}}$ can be described by
\begin{equation}
M_{\text{C}_\text{BA}}=M_{w_3}M_{h_2}M_{w_2}M_{h_1}M_{w_1}.
\end{equation}
Therefore, the transmission of the 1D-TPC can be expressed by
\begin{equation}
\begin{bmatrix}
b_2\\
a_2\end{bmatrix}=[M_\text{1D-TPC}]\begin{bmatrix}
a_1\\
b_1 \end{bmatrix}=[M_{\text{C}_\text{AB}}]^N[M_{\text{C}_\text{BA}}]^N\begin{bmatrix}
a_1\\
b_1 \end{bmatrix}.
\end{equation}
The transfer matrices $[M_\text{1D-TPC}]$ relates the incoming and outgoing wave amplitudes $a_{1,2}$ and $b_{1,2}$ at the same time. $N$ is the periods of the unit cells. Then the normalized amplitude transmission coefficient can be obtained from the relationship
\begin{equation}
t_{\text{1D-TPC}}=b_2,
\end{equation}
by assuming $a_1=1$ and $a_2=0$. For the whole system, the discrete TBS mode would interfere with the continuum leaky mode at sides of the bus-waveguide and results in the Fano resonance profile. Thus, the final transmission is
\begin{equation}\label{eq:eq5}
T=\left|A_\text{c}e^{i\phi}+A_\text{d}t_{\text{1D-TPC}}\right|^2,
\end{equation}
where $\phi$ is the relative phase between the two states and $A_\text{c}$ and $A_\text{d}$ are respectively amplitudes of them.

By assuming $r_1=0.43,~r_2=0.23$,$~\delta_1=\delta_3=0.775\pi/\lambda$[$\upmu$m] and $N=10$, for example, the transmission spectra can be obtained numerically from Eq.~\ref{eq:eq5}. An isolated high-$Q$ resonance peak at the wavelength of 1550 nm emerges in the center of the ultra-wide stopband background with a band range over 100 nm, as displayed in Fig.~\ref{fig:fig2}(b). This confirms the existence of TBS discussed in Fig.~\ref{fig:fig1}. In addition, as the reflectivity of the PTEs increases by, such as 10\% ($r_1=0.473,~r_2=0.253$) and 20\% ($r_1=0.516,~r_2=0.276$), the linewidth of the TBS peak narrows as well. However, the resonant wavelength remains unchanged though the reflectivity varies. This gives a special property to compare with a traditional PC nanobeam cavity based on Bragg reflection~\cite{ohta2011strong}. As shown in Fig.~\ref{fig:fig2}(c), 0th-order resonance mode of the Bragg-nanobeam is similarly calculated using transfer matrix at 1520 nm with a reference air-hole reflectivity $r=0.42$. When the reflectivity of the air-hole increases by 10\% ($r=0.462$) and 20\% ($r=0.504$), the resonant wavelength shifts with the reflectivity variation. It will cause a large deviation in the expected working wavelength during device fabrication. This robust feature of 1D-TPC contributes to a tolerant way to solve the problem of wavelength shift caused mode mis-overlapping between coupled cavities. For example, it is beneficial for the Fano resonance that arises from the interference between several discrete states~\cite{zangeneh2019topological,wang2020robust}.

In the proposed structure schematically shown in Fig.~\ref{fig:fig2}(a), we consider the interaction between the continuum high-order leaky mode and the discrete TBS with a relative amplitude ratio of, for example, $\alpha=A_\text{c}/A_\text{d}=0.2:0.8$. Then asymmetric Fano lineshapes can be obtained with the change of relative phase $\phi$. Figure~\ref{fig:fig2}(d) shows the calculated Fano lineshapes via different relative phase $\phi$. When $\phi=0$, standard Lorentzian lineshape can be observed and the asymmetric factor $q$ tends to infinity. For $\phi=\pm\pi/2$ by contrast, both constructive and destructive interferences occur. The interference between the TBS and leaky mode gives rise to the sharp asymmetric Fano lineshapes in transmission spectrum. With a doubled ER and a much higher slope rate (SR), this is beneficial for high sensitivity sensing and low power consumption optical switching. 

For a fixed relative phase of $\pi/2$, for example, the ER of the Fano lineshape can be further improved by adjusting the relative strength $\alpha$, as plotted in Fig.~\ref{fig:fig2}(e). Obviously, a decreased ratio of the continuum mode can cause a significant increase of ER while slightly sacrifice SR. However, it is not easy to change the parameters independently as the relative phase $\phi$ is closely related to the amplitude ratio $\alpha$ for a multi-mode system. This may contribute to the complex kinky phase-matching relationship during the mode conversion in the taper region. It is necessary to find a balance between these two points. 

\section{Experiment}
\begin{figure}
	\centering
	\includegraphics[width=\linewidth]{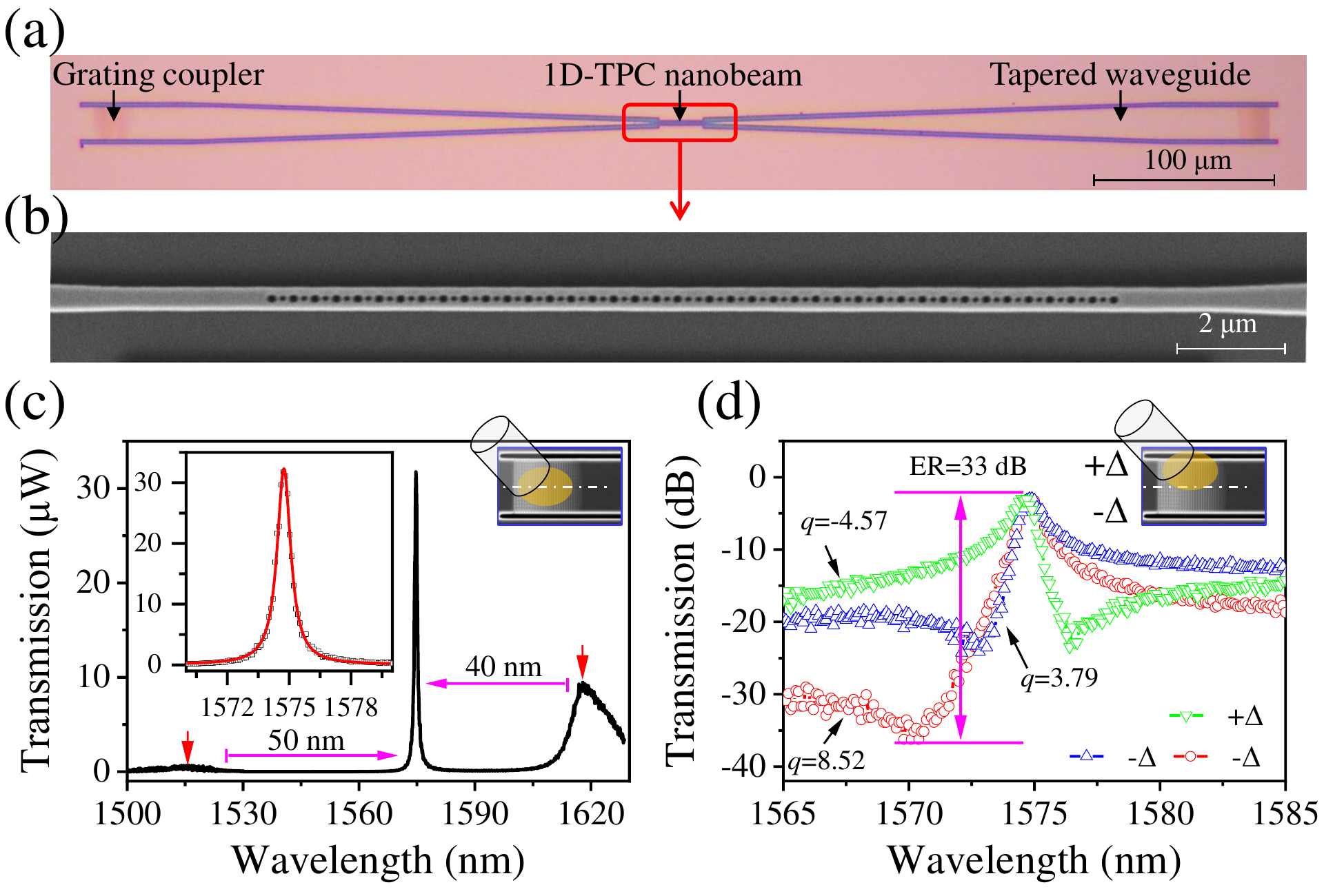}
	\caption{ (a) Optical micrograph of the fabricated device. (b) Zoomed SEM image of the 1D-TPC in (a). (c) Transmission spectrum of the fabricated device, showing a single isolated TBS at 1574.7 nm. Left inset is the fitted lineshape. Upper right inset gives the schematic image of the input fiber position, where the yellow ellipse indicates the surface spot of the fiber on the grating coupler. (d) Transmission spectra of the asymmetric Fano lineshapes obtained with different fiber dislocations. Upper right inset gives the schematic image of the input fiber position.}
	\label{fig:fig3}
\end{figure}

To verify above theoretical analyses, we fabricate several proposed 1D-TPCs with different structural parameters embedded in a bus-waveguide. The devices are fabricated on a commercial silicon-on-insulator (SOI) substrate with a 220 nm thick top silicon layer and a 2 $\upmu$m thick buried oxide layer. Electron beam lithography is used to define the device patterns followed by the transfer into the top silicon layer by inductively coupled plasma etching. Telecom-band grating couplers centered around 1575 nm with two-dimensional air-hole arrays are designed at both ends of the bus-waveguide~\cite{liu2010high}. The mode size converter is constructed with 200 $\upmu$m-long waveguide laterally tapered from 15 $\upmu$m to directly connected with a 500 nm wide bus-waveguide. Figure~\ref{fig:fig3}(a) displays an optical microscope image of the fabricated device. The scanning electron microscope (SEM) image of the 1D-TPC is zoomed in Fig.~\ref{fig:fig3}(b) with a compact footprint of 16 $\upmu$m $\times$ 0.5 $\upmu$m. 20 periods of unit cells are arranged on each side of the topological interface with an adjusted period of $\Lambda=390$ nm. The air-hole diameters are measured of 154 nm and 94 nm, respectively.

The fabricated devices are characterized by coupling a narrowband tunable laser (TL) into the input grating coupler, and the transmission powers are monitored by an optical power meter at the other grating coupler. By sweeping the TL wavelength over a range of 1500 to 1630 nm in steps of 0.1 nm, the device transmission spectrum could be obtained. As displayed in Fig.~\ref{fig:fig3}(c), an isolated TBS appears at wavelength of 1574.7 nm in a stopband background with a band range over 90 nm. The $Q$ factor is fitted to be 1,950 in the inset with standard Lorentzian function and this value could be much higher by further structure optimizations.

To realize the proposed asymmetric Fano resonance, an extra relative phase $\phi\neq0$ between the continuum leaky mode and the discrete TBS is needed, as discussed in Fig.~\ref{fig:fig2}(d). In our experiment, we achieve this relative phase by slightly shifting the input fiber with respect to the grating coupler~\cite{mehta2013fano}. The output fiber is fixed considering its location would not act any effect on the propagating modes. When the fiber is aligned to the central axis of the grating coupler (shown in the upper right inset of Fig.~\ref{fig:fig3}(c)), the relative phase $\phi$ is equal to zero owing to the internal symmetry. So that the resonant transmission of the 1D-TPC dominates to support discrete TBS as shown in Fig.~\ref{fig:fig3}(c) with an asymmetric factor $q$ as infinity. If the input fiber has an upward or downward dislocation $\Delta$ relative to the central axis of grating coupler, $\phi$ is no longer equal to zero due to the broken symmetry. A larger dislocation will result in a larger $\phi$ between the two states. A qualitative relationship between the phase and the dislocation can be summarized by comparing the theoretical analysis with the experimental data, which is $\phi\sim$$\Delta/\Delta_{\pi}$. $\Delta_{\pi}$ is the dislocation needed for a phase difference of $\pi$. This relationship is reported to be closely related to the opposite parity between the fundamental (even) and high-order modes (odd) that excited in the tapered waveguide~\cite{mehta2013fano}.

For the case that the fiber has an upward dislocation, as shown in the inset of Fig.~\ref{fig:fig3}(d), the relative phase $\phi$ is positive. Then the spectrum presents asymmetric Fano lineshape with an ER of 20.5 dB and a SR of 10.78 dB/nm, as shown by the green line in Fig.~\ref{fig:fig3}(d). The asymmetric factor $q$ is equal to -4.57 corresponding to a $\phi$ close to $\pi/2$ and an amplitude ratio $\alpha$ around 0.52:0.48 between the modes according to Eq.~\ref{eq:eq5}. However, for the case with a downward dislocated input fiber, the lineshape will become an inverted Fano lineshape (as presented in the red lines of Fig.~\ref{fig:fig3}(d)) with an negative relative phase around $-\pi/2$, which agrees with the theoretical analysis in Fig.~\ref{fig:fig2}(d). Parameters of the lines are calculated to have a similar ER of 21 dB and an SR of 10.07 dB/nm. A much higher ER is realized as discussed theoretically in Fig.~\ref{fig:fig2}(e) by carefully tuning the fiber dislocation to decrease the proportion of the continuum leaky mode to about 0.3. With a relevant $\phi$ of $-0.35\pi$, the measured ER is about 33 dB at the cost of a slight decrease in SR. This value is expected to be much higher than 33 dB theoretically ($\sim$45 dB), which is limited by the detection limit of the employed photodiode. Note, though the  relative phase $\phi$ is controlled by the fiber position here, 
optomechanics or integrated photonic MEMS are expected to realize a more precise and flexible dislocation in future device for more compact devices. Besides, compact mode multiplexer can be integrated as well to realize a similar operation\cite{zhang2018generating}.

\section{Conclusion}
In conclusion, we have demonstrated an ultra-compact 1D-TPC structure to generate Fano resonance lineshapes by interfering a discrete TBS with a continuum high-order leaky mode. With the topological feature of the demonstrated 1D-TPC, the TBS represents as an isolated resonance peak in the center of a flat background over 90 nm, which is beneficial for the single mode operation of the on-chip device applications. By exciting the continuum high-order leaky mode of the 1D-TPC, the Fano resonance lineshapes are obtained. From the fabricated device, lineshapes with an ER over 33 dB and a SR around 10 dB/nm are obtained, which could also be tuned by controlling the coupling condition of the input light.  Our results might open a new avenue to develop high-performance on-chip optical switch, sensing, and laser based on Fano resonance lineshapes. 

~\\
\section*{Fundings}\label{sec:05}
The Key Research and Development Program (2017YFA0303800), National Natural Science Foundation (91950119, 11634010, 61775183, 61905196), the Key Research and Development Program in Shaanxi Province of China (2020JZ-10), the Fundamental Research Funds for the Central Universities (310201911cx032, 3102019JC008).

~\\
\section*{Acknowledgment}\label{sec:06}
The authors would thank the Analytical \& Testing Center of NPU for the assistances of device fabrication.

\bibliographystyle{plain}

\begin{thebibliography}{38}%
	\makeatletter
	\providecommand \@ifxundefined [1]{%
		\@ifx{#1\undefined}
	}%
	\providecommand \@ifnum [1]{%
		\ifnum #1\expandafter \@firstoftwo
		\else \expandafter \@secondoftwo
		\fi
	}%
	\providecommand \@ifx [1]{%
		\ifx #1\expandafter \@firstoftwo
		\else \expandafter \@secondoftwo
		\fi
	}%
	\providecommand \natexlab [1]{#1}%
	\providecommand \enquote  [1]{``#1''}%
	\providecommand \bibnamefont  [1]{#1}%
	\providecommand \bibfnamefont [1]{#1}%
	\providecommand \citenamefont [1]{#1}%
	\providecommand \href@noop [0]{\@secondoftwo}%
	\providecommand \href [0]{\begingroup \@sanitize@url \@href}%
	\providecommand \@href[1]{\@@startlink{#1}\@@href}%
	\providecommand \@@href[1]{\endgroup#1\@@endlink}%
	\providecommand \@sanitize@url [0]{\catcode `\\12\catcode `\$12\catcode
		`\&12\catcode `\#12\catcode `\^12\catcode `\_12\catcode `\%12\relax}%
	\providecommand \@@startlink[1]{}%
	\providecommand \@@endlink[0]{}%
	\providecommand \url  [0]{\begingroup\@sanitize@url \@url }%
	\providecommand \@url [1]{\endgroup\@href {#1}{\urlprefix }}%
	\providecommand \urlprefix  [0]{URL }%
	\providecommand \Eprint [0]{\href }%
	\providecommand \doibase [0]{http://dx.doi.org/}%
	\providecommand \selectlanguage [0]{\@gobble}%
	\providecommand \bibinfo  [0]{\@secondoftwo}%
	\providecommand \bibfield  [0]{\@secondoftwo}%
	\providecommand \translation [1]{[#1]}%
	\providecommand \BibitemOpen [0]{}%
	\providecommand \bibitemStop [0]{}%
	\providecommand \bibitemNoStop [0]{.\EOS\space}%
	\providecommand \EOS [0]{\spacefactor3000\relax}%
	\providecommand \BibitemShut  [1]{\csname bibitem#1\endcsname}%
	\let\auto@bib@innerbib\@empty
	\bibitem [{\citenamefont {Fano}(1961)}]{fano1961effects}%
	\BibitemOpen
	\bibfield  {author} {\bibinfo {author} {\bibfnamefont {U.}~\bibnamefont
			{Fano}},\ }\bibfield  {title} {\enquote {\bibinfo {title} {Effects of
				configuration interaction on intensities and phase shifts},}\ }\href@noop {}
	{\bibfield  {journal} {\bibinfo  {journal} {Physical Review}\ }\textbf
		{\bibinfo {volume} {124}},\ \bibinfo {pages} {1866} (\bibinfo {year}
		{1961})}\BibitemShut {NoStop}%
	\bibitem [{\citenamefont {Yu}\ \emph {et~al.}(2016)\citenamefont {Yu},
		\citenamefont {Xue}, \citenamefont {Hu}, \citenamefont {Oxenl{\o}we},
		\citenamefont {Yvind},\ and\ \citenamefont {Mork}}]{yu2016all}%
	\BibitemOpen
	\bibfield  {author} {\bibinfo {author} {\bibfnamefont {Y.}~\bibnamefont
			{Yu}}, \bibinfo {author} {\bibfnamefont {W.}~\bibnamefont {Xue}}, \bibinfo
		{author} {\bibfnamefont {H.}~\bibnamefont {Hu}}, \bibinfo {author}
		{\bibfnamefont {L.~K.}\ \bibnamefont {Oxenl{\o}we}}, \bibinfo {author}
		{\bibfnamefont {K.}~\bibnamefont {Yvind}}, \ and\ \bibinfo {author}
		{\bibfnamefont {J.}~\bibnamefont {Mork}},\ }\bibfield  {title} {\enquote
		{\bibinfo {title} {All-optical switching improvement using photonic-crystal
				fano structures},}\ }\href@noop {} {\bibfield  {journal} {\bibinfo  {journal}
			{IEEE Photonics Journal}\ }\textbf {\bibinfo {volume} {8}},\ \bibinfo {pages}
		{1--8} (\bibinfo {year} {2016})}\BibitemShut {NoStop}%
	\bibitem [{\citenamefont {Limonov}\ \emph {et~al.}(2017)\citenamefont
		{Limonov}, \citenamefont {Rybin}, \citenamefont {Poddubny},\ and\
		\citenamefont {Kivshar}}]{limonov2017fano}%
	\BibitemOpen
	\bibfield  {author} {\bibinfo {author} {\bibfnamefont {M.~F.}\ \bibnamefont
			{Limonov}}, \bibinfo {author} {\bibfnamefont {M.~V.}\ \bibnamefont {Rybin}},
		\bibinfo {author} {\bibfnamefont {A.~N.}\ \bibnamefont {Poddubny}}, \ and\
		\bibinfo {author} {\bibfnamefont {Y.~S.}\ \bibnamefont {Kivshar}},\
	}\bibfield  {title} {\enquote {\bibinfo {title} {Fano resonances in
				photonics},}\ }\href@noop {} {\bibfield  {journal} {\bibinfo  {journal}
			{Nature Photonics}\ }\textbf {\bibinfo {volume} {11}},\ \bibinfo {pages}
		{543} (\bibinfo {year} {2017})}\BibitemShut {NoStop}%
	\bibitem [{\citenamefont {Yu}\ \emph {et~al.}(2017)\citenamefont {Yu},
		\citenamefont {Xue}, \citenamefont {Semenova}, \citenamefont {Yvind},\ and\
		\citenamefont {Mork}}]{yu2017demonstration}%
	\BibitemOpen
	\bibfield  {author} {\bibinfo {author} {\bibfnamefont {Y.}~\bibnamefont
			{Yu}}, \bibinfo {author} {\bibfnamefont {W.}~\bibnamefont {Xue}}, \bibinfo
		{author} {\bibfnamefont {E.}~\bibnamefont {Semenova}}, \bibinfo {author}
		{\bibfnamefont {K.}~\bibnamefont {Yvind}}, \ and\ \bibinfo {author}
		{\bibfnamefont {J.}~\bibnamefont {Mork}},\ }\bibfield  {title} {\enquote
		{\bibinfo {title} {Demonstration of a self-pulsing photonic crystal fano
				laser},}\ }\href@noop {} {\bibfield  {journal} {\bibinfo  {journal} {Nature
				Photonics}\ }\textbf {\bibinfo {volume} {11}},\ \bibinfo {pages} {81--84}
		(\bibinfo {year} {2017})}\BibitemShut {NoStop}%
	\bibitem [{\citenamefont {Yi}, \citenamefont {Citrin},\ and\ \citenamefont
		{Zhou}(2010)}]{yi2010highly}%
	\BibitemOpen
	\bibfield  {author} {\bibinfo {author} {\bibfnamefont {H.}~\bibnamefont
			{Yi}}, \bibinfo {author} {\bibfnamefont {D.}~\bibnamefont {Citrin}}, \ and\
		\bibinfo {author} {\bibfnamefont {Z.}~\bibnamefont {Zhou}},\ }\bibfield
	{title} {\enquote {\bibinfo {title} {Highly sensitive silicon microring
				sensor with sharp asymmetrical resonance},}\ }\href@noop {} {\bibfield
		{journal} {\bibinfo  {journal} {Optics Express}\ }\textbf {\bibinfo {volume}
			{18}},\ \bibinfo {pages} {2967--2972} (\bibinfo {year} {2010})}\BibitemShut
	{NoStop}%
	\bibitem [{\citenamefont {Tu}\ \emph {et~al.}(2017)\citenamefont {Tu},
		\citenamefont {Gao}, \citenamefont {Zhang},\ and\ \citenamefont
		{Zhang}}]{tu2017high}%
	\BibitemOpen
	\bibfield  {author} {\bibinfo {author} {\bibfnamefont {Z.}~\bibnamefont
			{Tu}}, \bibinfo {author} {\bibfnamefont {D.}~\bibnamefont {Gao}}, \bibinfo
		{author} {\bibfnamefont {M.}~\bibnamefont {Zhang}}, \ and\ \bibinfo {author}
		{\bibfnamefont {D.}~\bibnamefont {Zhang}},\ }\bibfield  {title} {\enquote
		{\bibinfo {title} {High-sensitivity complex refractive index sensing based on
				fano resonance in the subwavelength grating waveguide micro-ring
				resonator},}\ }\href@noop {} {\bibfield  {journal} {\bibinfo  {journal}
			{Optics Express}\ }\textbf {\bibinfo {volume} {25}},\ \bibinfo {pages}
		{20911--20922} (\bibinfo {year} {2017})}\BibitemShut {NoStop}%
	\bibitem [{\citenamefont {Zhang}\ \emph
		{et~al.}(2020{\natexlab{a}})\citenamefont {Zhang}, \citenamefont {Kang},
		\citenamefont {Xiong}, \citenamefont {Xu}, \citenamefont {Gu}, \citenamefont
		{Gan}, \citenamefont {Pan},\ and\ \citenamefont {Qu}}]{zhang2020photonic}%
	\BibitemOpen
	\bibfield  {author} {\bibinfo {author} {\bibfnamefont {C.}~\bibnamefont
			{Zhang}}, \bibinfo {author} {\bibfnamefont {G.}~\bibnamefont {Kang}},
		\bibinfo {author} {\bibfnamefont {Y.}~\bibnamefont {Xiong}}, \bibinfo
		{author} {\bibfnamefont {T.}~\bibnamefont {Xu}}, \bibinfo {author}
		{\bibfnamefont {L.}~\bibnamefont {Gu}}, \bibinfo {author} {\bibfnamefont
			{X.}~\bibnamefont {Gan}}, \bibinfo {author} {\bibfnamefont {Y.}~\bibnamefont
			{Pan}}, \ and\ \bibinfo {author} {\bibfnamefont {J.}~\bibnamefont {Qu}},\
	}\bibfield  {title} {\enquote {\bibinfo {title} {Photonic thermometer with a
				sub-millikelvin resolution and broad temperature range by waveguide-microring
				fano resonance},}\ }\href@noop {} {\bibfield  {journal} {\bibinfo  {journal}
			{Optics Express}\ }\textbf {\bibinfo {volume} {28}},\ \bibinfo {pages}
		{12599--12608} (\bibinfo {year} {2020}{\natexlab{a}})}\BibitemShut {NoStop}%
	\bibitem [{\citenamefont {Nozaki}\ \emph {et~al.}(2013)\citenamefont {Nozaki},
		\citenamefont {Shinya}, \citenamefont {Matsuo}, \citenamefont {Sato},
		\citenamefont {Kuramochi},\ and\ \citenamefont
		{Notomi}}]{nozaki2013ultralow}%
	\BibitemOpen
	\bibfield  {author} {\bibinfo {author} {\bibfnamefont {K.}~\bibnamefont
			{Nozaki}}, \bibinfo {author} {\bibfnamefont {A.}~\bibnamefont {Shinya}},
		\bibinfo {author} {\bibfnamefont {S.}~\bibnamefont {Matsuo}}, \bibinfo
		{author} {\bibfnamefont {T.}~\bibnamefont {Sato}}, \bibinfo {author}
		{\bibfnamefont {E.}~\bibnamefont {Kuramochi}}, \ and\ \bibinfo {author}
		{\bibfnamefont {M.}~\bibnamefont {Notomi}},\ }\bibfield  {title} {\enquote
		{\bibinfo {title} {Ultralow-energy and high-contrast all-optical switch
				involving fano resonance based on coupled photonic crystal nanocavities},}\
	}\href@noop {} {\bibfield  {journal} {\bibinfo  {journal} {Optics Express}\
		}\textbf {\bibinfo {volume} {21}},\ \bibinfo {pages} {11877--11888} (\bibinfo
		{year} {2013})}\BibitemShut {NoStop}%
	\bibitem [{\citenamefont {Yu}\ \emph {et~al.}(2014)\citenamefont {Yu},
		\citenamefont {Heuck}, \citenamefont {Hu}, \citenamefont {Xue}, \citenamefont
		{Peucheret}, \citenamefont {Chen}, \citenamefont {Oxenl{\o}we}, \citenamefont
		{Yvind},\ and\ \citenamefont {M{\o}rk}}]{yu2014fano}%
	\BibitemOpen
	\bibfield  {author} {\bibinfo {author} {\bibfnamefont {Y.}~\bibnamefont
			{Yu}}, \bibinfo {author} {\bibfnamefont {M.}~\bibnamefont {Heuck}}, \bibinfo
		{author} {\bibfnamefont {H.}~\bibnamefont {Hu}}, \bibinfo {author}
		{\bibfnamefont {W.}~\bibnamefont {Xue}}, \bibinfo {author} {\bibfnamefont
			{C.}~\bibnamefont {Peucheret}}, \bibinfo {author} {\bibfnamefont
			{Y.}~\bibnamefont {Chen}}, \bibinfo {author} {\bibfnamefont {L.~K.}\
			\bibnamefont {Oxenl{\o}we}}, \bibinfo {author} {\bibfnamefont
			{K.}~\bibnamefont {Yvind}}, \ and\ \bibinfo {author} {\bibfnamefont
			{J.}~\bibnamefont {M{\o}rk}},\ }\bibfield  {title} {\enquote {\bibinfo
			{title} {Fano resonance control in a photonic crystal structure and its
				application to ultrafast switching},}\ }\href@noop {} {\bibfield  {journal}
		{\bibinfo  {journal} {Applied Physics Letters}\ }\textbf {\bibinfo {volume}
			{105}},\ \bibinfo {pages} {061117} (\bibinfo {year} {2014})}\BibitemShut
	{NoStop}%
	\bibitem [{\citenamefont {Heuck}\ \emph {et~al.}(2013)\citenamefont {Heuck},
		\citenamefont {Kristensen}, \citenamefont {Elesin},\ and\ \citenamefont
		{M{\o}rk}}]{heuck2013improved}%
	\BibitemOpen
	\bibfield  {author} {\bibinfo {author} {\bibfnamefont {M.}~\bibnamefont
			{Heuck}}, \bibinfo {author} {\bibfnamefont {P.~T.}\ \bibnamefont
			{Kristensen}}, \bibinfo {author} {\bibfnamefont {Y.}~\bibnamefont {Elesin}},
		\ and\ \bibinfo {author} {\bibfnamefont {J.}~\bibnamefont {M{\o}rk}},\
	}\bibfield  {title} {\enquote {\bibinfo {title} {Improved switching using
				fano resonances in photonic crystal structures},}\ }\href@noop {} {\bibfield
		{journal} {\bibinfo  {journal} {Optics Letters}\ }\textbf {\bibinfo {volume}
			{38}},\ \bibinfo {pages} {2466--2468} (\bibinfo {year} {2013})}\BibitemShut
	{NoStop}%
	\bibitem [{\citenamefont {Zhang}, \citenamefont {Li},\ and\ \citenamefont
		{Yao}(2016)}]{zhang2016optically}%
	\BibitemOpen
	\bibfield  {author} {\bibinfo {author} {\bibfnamefont {W.}~\bibnamefont
			{Zhang}}, \bibinfo {author} {\bibfnamefont {W.}~\bibnamefont {Li}}, \ and\
		\bibinfo {author} {\bibfnamefont {J.}~\bibnamefont {Yao}},\ }\bibfield
	{title} {\enquote {\bibinfo {title} {Optically tunable fano resonance in a
				grating-based fabry--perot cavity-coupled microring resonator on a silicon
				chip},}\ }\href@noop {} {\bibfield  {journal} {\bibinfo  {journal} {Optics
				Letters}\ }\textbf {\bibinfo {volume} {41}},\ \bibinfo {pages} {2474--2477}
		(\bibinfo {year} {2016})}\BibitemShut {NoStop}%
	\bibitem [{\citenamefont {Qiu}\ \emph {et~al.}(2012)\citenamefont {Qiu},
		\citenamefont {Yu}, \citenamefont {Hu}, \citenamefont {Wang}, \citenamefont
		{Jiang},\ and\ \citenamefont {Yang}}]{qiu2012asymmetric}%
	\BibitemOpen
	\bibfield  {author} {\bibinfo {author} {\bibfnamefont {C.}~\bibnamefont
			{Qiu}}, \bibinfo {author} {\bibfnamefont {P.}~\bibnamefont {Yu}}, \bibinfo
		{author} {\bibfnamefont {T.}~\bibnamefont {Hu}}, \bibinfo {author}
		{\bibfnamefont {F.}~\bibnamefont {Wang}}, \bibinfo {author} {\bibfnamefont
			{X.}~\bibnamefont {Jiang}}, \ and\ \bibinfo {author} {\bibfnamefont
			{J.}~\bibnamefont {Yang}},\ }\bibfield  {title} {\enquote {\bibinfo {title}
			{Asymmetric fano resonance in eye-like microring system},}\ }\href@noop {}
	{\bibfield  {journal} {\bibinfo  {journal} {Applied Physics Letters}\
		}\textbf {\bibinfo {volume} {101}},\ \bibinfo {pages} {021110} (\bibinfo
		{year} {2012})}\BibitemShut {NoStop}%
	\bibitem [{\citenamefont {Gu}\ \emph {et~al.}(2019)\citenamefont {Gu},
		\citenamefont {Fang}, \citenamefont {Li}, \citenamefont {Fang}, \citenamefont
		{Chua}, \citenamefont {Zhao},\ and\ \citenamefont {Gan}}]{gu2019compact}%
	\BibitemOpen
	\bibfield  {author} {\bibinfo {author} {\bibfnamefont {L.}~\bibnamefont
			{Gu}}, \bibinfo {author} {\bibfnamefont {H.}~\bibnamefont {Fang}}, \bibinfo
		{author} {\bibfnamefont {J.}~\bibnamefont {Li}}, \bibinfo {author}
		{\bibfnamefont {L.}~\bibnamefont {Fang}}, \bibinfo {author} {\bibfnamefont
			{S.~J.}\ \bibnamefont {Chua}}, \bibinfo {author} {\bibfnamefont
			{J.}~\bibnamefont {Zhao}}, \ and\ \bibinfo {author} {\bibfnamefont
			{X.}~\bibnamefont {Gan}},\ }\bibfield  {title} {\enquote {\bibinfo {title} {A
				compact structure for realizing lorentzian, fano, and electromagnetically
				induced transparency resonance lineshapes in a microring resonator},}\
	}\href@noop {} {\bibfield  {journal} {\bibinfo  {journal} {Nanophotonics}\
		}\textbf {\bibinfo {volume} {8}},\ \bibinfo {pages} {841--848} (\bibinfo
		{year} {2019})}\BibitemShut {NoStop}%
	\bibitem [{\citenamefont {Gu}\ \emph {et~al.}(2020)\citenamefont {Gu},
		\citenamefont {Fang}, \citenamefont {Fang}, \citenamefont {Li}, \citenamefont
		{Zheng}, \citenamefont {Zhao}, \citenamefont {Zhao},\ and\ \citenamefont
		{Gan}}]{gu2020fano}%
	\BibitemOpen
	\bibfield  {author} {\bibinfo {author} {\bibfnamefont {L.}~\bibnamefont
			{Gu}}, \bibinfo {author} {\bibfnamefont {L.}~\bibnamefont {Fang}}, \bibinfo
		{author} {\bibfnamefont {H.}~\bibnamefont {Fang}}, \bibinfo {author}
		{\bibfnamefont {J.}~\bibnamefont {Li}}, \bibinfo {author} {\bibfnamefont
			{J.}~\bibnamefont {Zheng}}, \bibinfo {author} {\bibfnamefont
			{J.}~\bibnamefont {Zhao}}, \bibinfo {author} {\bibfnamefont {Q.}~\bibnamefont
			{Zhao}}, \ and\ \bibinfo {author} {\bibfnamefont {X.}~\bibnamefont {Gan}},\
	}\bibfield  {title} {\enquote {\bibinfo {title} {Fano resonance lineshapes in
				a waveguide-microring structure enabled by an air-hole},}\ }\href@noop {}
	{\bibfield  {journal} {\bibinfo  {journal} {APL Photonics}\ }\textbf
		{\bibinfo {volume} {5}},\ \bibinfo {pages} {016108} (\bibinfo {year}
		{2020})}\BibitemShut {NoStop}%
	\bibitem [{\citenamefont {Fang}\ \emph {et~al.}(2020)\citenamefont {Fang},
		\citenamefont {Gu}, \citenamefont {Zheng}, \citenamefont {Zhao},
		\citenamefont {Gan},\ and\ \citenamefont {Zhao}}]{fang2020controlling}%
	\BibitemOpen
	\bibfield  {author} {\bibinfo {author} {\bibfnamefont {L.}~\bibnamefont
			{Fang}}, \bibinfo {author} {\bibfnamefont {L.}~\bibnamefont {Gu}}, \bibinfo
		{author} {\bibfnamefont {J.}~\bibnamefont {Zheng}}, \bibinfo {author}
		{\bibfnamefont {Q.}~\bibnamefont {Zhao}}, \bibinfo {author} {\bibfnamefont
			{X.}~\bibnamefont {Gan}}, \ and\ \bibinfo {author} {\bibfnamefont
			{J.}~\bibnamefont {Zhao}},\ }\bibfield  {title} {\enquote {\bibinfo {title}
			{Controlling resonance lineshapes of a side-coupled waveguide-microring
				resonator},}\ }\href@noop {} {\bibfield  {journal} {\bibinfo  {journal}
			{Journal of Lightwave Technology}\ }\textbf {\bibinfo {volume} {38}},\
		\bibinfo {pages} {4429--4434} (\bibinfo {year} {2020})}\BibitemShut {NoStop}%
	\bibitem [{\citenamefont {Li}\ \emph {et~al.}(2012)\citenamefont {Li},
		\citenamefont {Xiao}, \citenamefont {Zou}, \citenamefont {Jiang},
		\citenamefont {Liu}, \citenamefont {Sun}, \citenamefont {Li},\ and\
		\citenamefont {Gong}}]{li2012experimental}%
	\BibitemOpen
	\bibfield  {author} {\bibinfo {author} {\bibfnamefont {B.-B.}\ \bibnamefont
			{Li}}, \bibinfo {author} {\bibfnamefont {Y.-F.}\ \bibnamefont {Xiao}},
		\bibinfo {author} {\bibfnamefont {C.-L.}\ \bibnamefont {Zou}}, \bibinfo
		{author} {\bibfnamefont {X.-F.}\ \bibnamefont {Jiang}}, \bibinfo {author}
		{\bibfnamefont {Y.-C.}\ \bibnamefont {Liu}}, \bibinfo {author} {\bibfnamefont
			{F.-W.}\ \bibnamefont {Sun}}, \bibinfo {author} {\bibfnamefont
			{Y.}~\bibnamefont {Li}}, \ and\ \bibinfo {author} {\bibfnamefont
			{Q.}~\bibnamefont {Gong}},\ }\bibfield  {title} {\enquote {\bibinfo {title}
			{Experimental controlling of fano resonance in indirectly coupled
				whispering-gallery microresonators},}\ }\href@noop {} {\bibfield  {journal}
		{\bibinfo  {journal} {Applied Physics Letters}\ }\textbf {\bibinfo {volume}
			{100}},\ \bibinfo {pages} {021108} (\bibinfo {year} {2012})}\BibitemShut
	{NoStop}%
	\bibitem [{\citenamefont {Raghu}\ and\ \citenamefont
		{Haldane}(2008)}]{raghu2008analogs}%
	\BibitemOpen
	\bibfield  {author} {\bibinfo {author} {\bibfnamefont {S.}~\bibnamefont
			{Raghu}}\ and\ \bibinfo {author} {\bibfnamefont {F.~D.~M.}\ \bibnamefont
			{Haldane}},\ }\bibfield  {title} {\enquote {\bibinfo {title} {Analogs of
				quantum-hall-effect edge states in photonic crystals},}\ }\href@noop {}
	{\bibfield  {journal} {\bibinfo  {journal} {Physical Review A}\ }\textbf
		{\bibinfo {volume} {78}},\ \bibinfo {pages} {033834} (\bibinfo {year}
		{2008})}\BibitemShut {NoStop}%
	\bibitem [{\citenamefont {Chen}\ \emph {et~al.}(2017)\citenamefont {Chen},
		\citenamefont {Zhao}, \citenamefont {Chen},\ and\ \citenamefont
		{Dong}}]{chen2017valley}%
	\BibitemOpen
	\bibfield  {author} {\bibinfo {author} {\bibfnamefont {X.-D.}\ \bibnamefont
			{Chen}}, \bibinfo {author} {\bibfnamefont {F.-L.}\ \bibnamefont {Zhao}},
		\bibinfo {author} {\bibfnamefont {M.}~\bibnamefont {Chen}}, \ and\ \bibinfo
		{author} {\bibfnamefont {J.-W.}\ \bibnamefont {Dong}},\ }\bibfield  {title}
	{\enquote {\bibinfo {title} {Valley-contrasting physics in all-dielectric
				photonic crystals: Orbital angular momentum and topological propagation},}\
	}\href@noop {} {\bibfield  {journal} {\bibinfo  {journal} {Physical Review
				B}\ }\textbf {\bibinfo {volume} {96}},\ \bibinfo {pages} {020202} (\bibinfo
		{year} {2017})}\BibitemShut {NoStop}%
	\bibitem [{\citenamefont {Khanikaev}\ and\ \citenamefont
		{Shvets}(2017)}]{khanikaev2017two}%
	\BibitemOpen
	\bibfield  {author} {\bibinfo {author} {\bibfnamefont {A.~B.}\ \bibnamefont
			{Khanikaev}}\ and\ \bibinfo {author} {\bibfnamefont {G.}~\bibnamefont
			{Shvets}},\ }\bibfield  {title} {\enquote {\bibinfo {title} {Two-dimensional
				topological photonics},}\ }\href@noop {} {\bibfield  {journal} {\bibinfo
			{journal} {Nature Photonics}\ }\textbf {\bibinfo {volume} {11}},\ \bibinfo
		{pages} {763--773} (\bibinfo {year} {2017})}\BibitemShut {NoStop}%
	\bibitem [{\citenamefont {Ota}\ \emph {et~al.}(2018)\citenamefont {Ota},
		\citenamefont {Katsumi}, \citenamefont {Watanabe}, \citenamefont {Iwamoto},\
		and\ \citenamefont {Arakawa}}]{ota2018topological}%
	\BibitemOpen
	\bibfield  {author} {\bibinfo {author} {\bibfnamefont {Y.}~\bibnamefont
			{Ota}}, \bibinfo {author} {\bibfnamefont {R.}~\bibnamefont {Katsumi}},
		\bibinfo {author} {\bibfnamefont {K.}~\bibnamefont {Watanabe}}, \bibinfo
		{author} {\bibfnamefont {S.}~\bibnamefont {Iwamoto}}, \ and\ \bibinfo
		{author} {\bibfnamefont {Y.}~\bibnamefont {Arakawa}},\ }\bibfield  {title}
	{\enquote {\bibinfo {title} {Topological photonic crystal nanocavity
				laser},}\ }\href@noop {} {\bibfield  {journal} {\bibinfo  {journal}
			{Communications Physics}\ }\textbf {\bibinfo {volume} {1}},\ \bibinfo {pages}
		{1--8} (\bibinfo {year} {2018})}\BibitemShut {NoStop}%
	\bibitem [{\citenamefont {Zhang}\ \emph
		{et~al.}(2020{\natexlab{b}})\citenamefont {Zhang}, \citenamefont {Xie},
		\citenamefont {Hao}, \citenamefont {Dang}, \citenamefont {Xiao},
		\citenamefont {Shi}, \citenamefont {Ni}, \citenamefont {Niu}, \citenamefont
		{Wang}, \citenamefont {Jin} \emph {et~al.}}]{zhang2020low}%
	\BibitemOpen
	\bibfield  {author} {\bibinfo {author} {\bibfnamefont {W.}~\bibnamefont
			{Zhang}}, \bibinfo {author} {\bibfnamefont {X.}~\bibnamefont {Xie}}, \bibinfo
		{author} {\bibfnamefont {H.}~\bibnamefont {Hao}}, \bibinfo {author}
		{\bibfnamefont {J.}~\bibnamefont {Dang}}, \bibinfo {author} {\bibfnamefont
			{S.}~\bibnamefont {Xiao}}, \bibinfo {author} {\bibfnamefont {S.}~\bibnamefont
			{Shi}}, \bibinfo {author} {\bibfnamefont {H.}~\bibnamefont {Ni}}, \bibinfo
		{author} {\bibfnamefont {Z.}~\bibnamefont {Niu}}, \bibinfo {author}
		{\bibfnamefont {C.}~\bibnamefont {Wang}}, \bibinfo {author} {\bibfnamefont
			{K.}~\bibnamefont {Jin}},  \emph {et~al.},\ }\bibfield  {title} {\enquote
		{\bibinfo {title} {Low-threshold topological nanolasers based on the
				second-order corner state},}\ }\href@noop {} {\bibfield  {journal} {\bibinfo
			{journal} {Light: Science \& Applications}\ }\textbf {\bibinfo {volume}
			{9}},\ \bibinfo {pages} {1--6} (\bibinfo {year}
		{2020}{\natexlab{b}})}\BibitemShut {NoStop}%
	\bibitem [{\citenamefont {Shao}\ \emph {et~al.}(2020)\citenamefont {Shao},
		\citenamefont {Chen}, \citenamefont {Wang}, \citenamefont {Mao},
		\citenamefont {Yang}, \citenamefont {Wang}, \citenamefont {Wang},
		\citenamefont {Hu},\ and\ \citenamefont {Ma}}]{shao2020high}%
	\BibitemOpen
	\bibfield  {author} {\bibinfo {author} {\bibfnamefont {Z.-K.}\ \bibnamefont
			{Shao}}, \bibinfo {author} {\bibfnamefont {H.-Z.}\ \bibnamefont {Chen}},
		\bibinfo {author} {\bibfnamefont {S.}~\bibnamefont {Wang}}, \bibinfo {author}
		{\bibfnamefont {X.-R.}\ \bibnamefont {Mao}}, \bibinfo {author} {\bibfnamefont
			{Z.-Q.}\ \bibnamefont {Yang}}, \bibinfo {author} {\bibfnamefont {S.-L.}\
			\bibnamefont {Wang}}, \bibinfo {author} {\bibfnamefont {X.-X.}\ \bibnamefont
			{Wang}}, \bibinfo {author} {\bibfnamefont {X.}~\bibnamefont {Hu}}, \ and\
		\bibinfo {author} {\bibfnamefont {R.-M.}\ \bibnamefont {Ma}},\ }\bibfield
	{title} {\enquote {\bibinfo {title} {A high-performance topological bulk
				laser based on band-inversion-induced reflection},}\ }\href@noop {}
	{\bibfield  {journal} {\bibinfo  {journal} {Nature Nanotechnology}\ }\textbf
		{\bibinfo {volume} {15}},\ \bibinfo {pages} {67--72} (\bibinfo {year}
		{2020})}\BibitemShut {NoStop}%
	\bibitem [{\citenamefont {Han}\ \emph {et~al.}(2019)\citenamefont {Han},
		\citenamefont {Lee}, \citenamefont {Callard}, \citenamefont {Seassal},\ and\
		\citenamefont {Jeon}}]{han2019lasing}%
	\BibitemOpen
	\bibfield  {author} {\bibinfo {author} {\bibfnamefont {C.}~\bibnamefont
			{Han}}, \bibinfo {author} {\bibfnamefont {M.}~\bibnamefont {Lee}}, \bibinfo
		{author} {\bibfnamefont {S.}~\bibnamefont {Callard}}, \bibinfo {author}
		{\bibfnamefont {C.}~\bibnamefont {Seassal}}, \ and\ \bibinfo {author}
		{\bibfnamefont {H.}~\bibnamefont {Jeon}},\ }\bibfield  {title} {\enquote
		{\bibinfo {title} {Lasing at topological edge states in a photonic crystal l3
				nanocavity dimer array},}\ }\href@noop {} {\bibfield  {journal} {\bibinfo
			{journal} {Light: Science \& Applications}\ }\textbf {\bibinfo {volume}
			{8}},\ \bibinfo {pages} {1--10} (\bibinfo {year} {2019})}\BibitemShut
	{NoStop}%
	\bibitem [{\citenamefont {Li}\ \emph {et~al.}(2018)\citenamefont {Li},
		\citenamefont {Wang}, \citenamefont {Xiong}, \citenamefont {Lou},
		\citenamefont {Chen}, \citenamefont {Wu}, \citenamefont {Poo}, \citenamefont
		{Jiang},\ and\ \citenamefont {John}}]{li2018topological}%
	\BibitemOpen
	\bibfield  {author} {\bibinfo {author} {\bibfnamefont {F.-F.}\ \bibnamefont
			{Li}}, \bibinfo {author} {\bibfnamefont {H.-X.}\ \bibnamefont {Wang}},
		\bibinfo {author} {\bibfnamefont {Z.}~\bibnamefont {Xiong}}, \bibinfo
		{author} {\bibfnamefont {Q.}~\bibnamefont {Lou}}, \bibinfo {author}
		{\bibfnamefont {P.}~\bibnamefont {Chen}}, \bibinfo {author} {\bibfnamefont
			{R.-X.}\ \bibnamefont {Wu}}, \bibinfo {author} {\bibfnamefont
			{Y.}~\bibnamefont {Poo}}, \bibinfo {author} {\bibfnamefont {J.-H.}\
			\bibnamefont {Jiang}}, \ and\ \bibinfo {author} {\bibfnamefont
			{S.}~\bibnamefont {John}},\ }\bibfield  {title} {\enquote {\bibinfo {title}
			{Topological light-trapping on a dislocation},}\ }\href@noop {} {\bibfield
		{journal} {\bibinfo  {journal} {Nature Communications}\ }\textbf {\bibinfo
			{volume} {9}},\ \bibinfo {pages} {1--8} (\bibinfo {year} {2018})}\BibitemShut
	{NoStop}%
	\bibitem [{\citenamefont {Shalaev}\ \emph {et~al.}(2019)\citenamefont
		{Shalaev}, \citenamefont {Walasik}, \citenamefont {Tsukernik}, \citenamefont
		{Xu},\ and\ \citenamefont {Litchinitser}}]{shalaev2019robust}%
	\BibitemOpen
	\bibfield  {author} {\bibinfo {author} {\bibfnamefont {M.~I.}\ \bibnamefont
			{Shalaev}}, \bibinfo {author} {\bibfnamefont {W.}~\bibnamefont {Walasik}},
		\bibinfo {author} {\bibfnamefont {A.}~\bibnamefont {Tsukernik}}, \bibinfo
		{author} {\bibfnamefont {Y.}~\bibnamefont {Xu}}, \ and\ \bibinfo {author}
		{\bibfnamefont {N.~M.}\ \bibnamefont {Litchinitser}},\ }\bibfield  {title}
	{\enquote {\bibinfo {title} {Robust topologically protected transport in
				photonic crystals at telecommunication wavelengths},}\ }\href@noop {}
	{\bibfield  {journal} {\bibinfo  {journal} {Nature Nanotechnology}\ }\textbf
		{\bibinfo {volume} {14}},\ \bibinfo {pages} {31--34} (\bibinfo {year}
		{2019})}\BibitemShut {NoStop}%
	\bibitem [{\citenamefont {He}\ \emph {et~al.}(2019)\citenamefont {He},
		\citenamefont {Liang}, \citenamefont {Yuan}, \citenamefont {Qiu},
		\citenamefont {Chen}, \citenamefont {Zhao},\ and\ \citenamefont
		{Dong}}]{he2019silicon}%
	\BibitemOpen
	\bibfield  {author} {\bibinfo {author} {\bibfnamefont {X.-T.}\ \bibnamefont
			{He}}, \bibinfo {author} {\bibfnamefont {E.-T.}\ \bibnamefont {Liang}},
		\bibinfo {author} {\bibfnamefont {J.-J.}\ \bibnamefont {Yuan}}, \bibinfo
		{author} {\bibfnamefont {H.-Y.}\ \bibnamefont {Qiu}}, \bibinfo {author}
		{\bibfnamefont {X.-D.}\ \bibnamefont {Chen}}, \bibinfo {author}
		{\bibfnamefont {F.-L.}\ \bibnamefont {Zhao}}, \ and\ \bibinfo {author}
		{\bibfnamefont {J.-W.}\ \bibnamefont {Dong}},\ }\bibfield  {title} {\enquote
		{\bibinfo {title} {A silicon-on-insulator slab for topological valley
				transport},}\ }\href@noop {} {\bibfield  {journal} {\bibinfo  {journal}
			{Nature Communications}\ }\textbf {\bibinfo {volume} {10}},\ \bibinfo {pages}
		{1--9} (\bibinfo {year} {2019})}\BibitemShut {NoStop}%
	\bibitem [{\citenamefont {Gao}\ \emph {et~al.}(2018)\citenamefont {Gao},
		\citenamefont {Hu}, \citenamefont {Li}, \citenamefont {Yang}, \citenamefont
		{Chai}, \citenamefont {Xie},\ and\ \citenamefont {Gong}}]{gao2018fano}%
	\BibitemOpen
	\bibfield  {author} {\bibinfo {author} {\bibfnamefont {W.}~\bibnamefont
			{Gao}}, \bibinfo {author} {\bibfnamefont {X.}~\bibnamefont {Hu}}, \bibinfo
		{author} {\bibfnamefont {C.}~\bibnamefont {Li}}, \bibinfo {author}
		{\bibfnamefont {J.}~\bibnamefont {Yang}}, \bibinfo {author} {\bibfnamefont
			{Z.}~\bibnamefont {Chai}}, \bibinfo {author} {\bibfnamefont {J.}~\bibnamefont
			{Xie}}, \ and\ \bibinfo {author} {\bibfnamefont {Q.}~\bibnamefont {Gong}},\
	}\bibfield  {title} {\enquote {\bibinfo {title} {Fano-resonance in
				one-dimensional topological photonic crystal heterostructure},}\ }\href@noop
	{} {\bibfield  {journal} {\bibinfo  {journal} {Optics Express}\ }\textbf
		{\bibinfo {volume} {26}},\ \bibinfo {pages} {8634--8644} (\bibinfo {year}
		{2018})}\BibitemShut {NoStop}%
	\bibitem [{\citenamefont {Zangeneh-Nejad}\ and\ \citenamefont
		{Fleury}(2019)}]{zangeneh2019topological}%
	\BibitemOpen
	\bibfield  {author} {\bibinfo {author} {\bibfnamefont {F.}~\bibnamefont
			{Zangeneh-Nejad}}\ and\ \bibinfo {author} {\bibfnamefont {R.}~\bibnamefont
			{Fleury}},\ }\bibfield  {title} {\enquote {\bibinfo {title} {Topological fano
				resonances},}\ }\href@noop {} {\bibfield  {journal} {\bibinfo  {journal}
			{Physical Review Letters}\ }\textbf {\bibinfo {volume} {122}},\ \bibinfo
		{pages} {014301} (\bibinfo {year} {2019})}\BibitemShut {NoStop}%
	\bibitem [{\citenamefont {Wang}\ \emph {et~al.}(2020)\citenamefont {Wang},
		\citenamefont {Jin}, \citenamefont {Wang}, \citenamefont {Bonello},
		\citenamefont {Djafari-Rouhani},\ and\ \citenamefont
		{Fleury}}]{wang2020robust}%
	\BibitemOpen
	\bibfield  {author} {\bibinfo {author} {\bibfnamefont {W.}~\bibnamefont
			{Wang}}, \bibinfo {author} {\bibfnamefont {Y.}~\bibnamefont {Jin}}, \bibinfo
		{author} {\bibfnamefont {W.}~\bibnamefont {Wang}}, \bibinfo {author}
		{\bibfnamefont {B.}~\bibnamefont {Bonello}}, \bibinfo {author} {\bibfnamefont
			{B.}~\bibnamefont {Djafari-Rouhani}}, \ and\ \bibinfo {author} {\bibfnamefont
			{R.}~\bibnamefont {Fleury}},\ }\bibfield  {title} {\enquote {\bibinfo {title}
			{Robust fano resonance in a topological mechanical beam},}\ }\href@noop {}
	{\bibfield  {journal} {\bibinfo  {journal} {Physical Review B}\ }\textbf
		{\bibinfo {volume} {101}},\ \bibinfo {pages} {024101} (\bibinfo {year}
		{2020})}\BibitemShut {NoStop}%
	\bibitem [{\citenamefont {Ji}, \citenamefont {Zhang},\ and\ \citenamefont
		{Yao}(2021)}]{ji2020topologically}%
	\BibitemOpen
	\bibfield  {author} {\bibinfo {author} {\bibfnamefont {C.-Y.}\ \bibnamefont
			{Ji}}, \bibinfo {author} {\bibfnamefont {Y.}~\bibnamefont {Zhang}}, \ and\
		\bibinfo {author} {\bibfnamefont {Y.}~\bibnamefont {Yao}},\ }\bibfield
	{title} {\enquote {\bibinfo {title} {Topologically protected fano resonance
				in photonic valley hall insulators},}\ }\href@noop {} {\bibfield  {journal}
		{\bibinfo  {journal} {Physical Review A}\ }\textbf {\bibinfo {volume}
			{103}},\ \bibinfo {pages} {023512} (\bibinfo {year} {2021})}\BibitemShut
	{NoStop}%
	\bibitem [{\citenamefont {Su}, \citenamefont {Schrieffer},\ and\ \citenamefont
		{Heeger}(1979)}]{su1979solitons}%
	\BibitemOpen
	\bibfield  {author} {\bibinfo {author} {\bibfnamefont {W.}~\bibnamefont
			{Su}}, \bibinfo {author} {\bibfnamefont {J.}~\bibnamefont {Schrieffer}}, \
		and\ \bibinfo {author} {\bibfnamefont {A.~J.}\ \bibnamefont {Heeger}},\
	}\bibfield  {title} {\enquote {\bibinfo {title} {Solitons in
				polyacetylene},}\ }\href@noop {} {\bibfield  {journal} {\bibinfo  {journal}
			{Physical Review Letters}\ }\textbf {\bibinfo {volume} {42}},\ \bibinfo
		{pages} {1698} (\bibinfo {year} {1979})}\BibitemShut {NoStop}%
	\bibitem [{\citenamefont {Xiao}, \citenamefont {Zhang},\ and\ \citenamefont
		{Chan}(2014)}]{xiao2014surface}%
	\BibitemOpen
	\bibfield  {author} {\bibinfo {author} {\bibfnamefont {M.}~\bibnamefont
			{Xiao}}, \bibinfo {author} {\bibfnamefont {Z.}~\bibnamefont {Zhang}}, \ and\
		\bibinfo {author} {\bibfnamefont {C.~T.}\ \bibnamefont {Chan}},\ }\bibfield
	{title} {\enquote {\bibinfo {title} {Surface impedance and bulk band
				geometric phases in one-dimensional systems},}\ }\href@noop {} {\bibfield
		{journal} {\bibinfo  {journal} {Physical Review X}\ }\textbf {\bibinfo
			{volume} {4}},\ \bibinfo {pages} {021017} (\bibinfo {year}
		{2014})}\BibitemShut {NoStop}%
	\bibitem [{\citenamefont {Joannopoulos}\ \emph {et~al.}(2008)\citenamefont
		{Joannopoulos}, \citenamefont {Johnson}, \citenamefont {Winn},\ and\
		\citenamefont {Meade}}]{joannopoulos2008molding}%
	\BibitemOpen
	\bibfield  {author} {\bibinfo {author} {\bibfnamefont {J.~D.}\ \bibnamefont
			{Joannopoulos}}, \bibinfo {author} {\bibfnamefont {S.~G.}\ \bibnamefont
			{Johnson}}, \bibinfo {author} {\bibfnamefont {J.~N.}\ \bibnamefont {Winn}}, \
		and\ \bibinfo {author} {\bibfnamefont {R.~D.}\ \bibnamefont {Meade}},\
	}\bibfield  {title} {\enquote {\bibinfo {title} {Molding the flow of
				light},}\ }\href@noop {} {\bibfield  {journal} {\bibinfo  {journal}
			{Princeton Univ. Press, Princeton, NJ [ua]}\ } (\bibinfo {year}
		{2008})}\BibitemShut {NoStop}%
	\bibitem [{\citenamefont {Xie}\ \emph {et~al.}(2021)\citenamefont {Xie},
		\citenamefont {Verheyen}, \citenamefont {Pantouvaki}, \citenamefont
		{Van~Campenhout},\ and\ \citenamefont {Van~Thourhout}}]{xie2021efficient}%
	\BibitemOpen
	\bibfield  {author} {\bibinfo {author} {\bibfnamefont {W.}~\bibnamefont
			{Xie}}, \bibinfo {author} {\bibfnamefont {P.}~\bibnamefont {Verheyen}},
		\bibinfo {author} {\bibfnamefont {M.}~\bibnamefont {Pantouvaki}}, \bibinfo
		{author} {\bibfnamefont {J.}~\bibnamefont {Van~Campenhout}}, \ and\ \bibinfo
		{author} {\bibfnamefont {D.}~\bibnamefont {Van~Thourhout}},\ }\bibfield
	{title} {\enquote {\bibinfo {title} {Efficient resonance management in
				ultrahigh-q 1d photonic crystal nanocavities fabricated on 300 mm soi cmos
				platform},}\ }\href@noop {} {\bibfield  {journal} {\bibinfo  {journal} {Laser
				\& Photonics Reviews}\ }\textbf {\bibinfo {volume} {15}},\ \bibinfo {pages}
		{2000317} (\bibinfo {year} {2021})}\BibitemShut {NoStop}%
	\bibitem [{\citenamefont {Mehta}, \citenamefont {Orcutt},\ and\ \citenamefont
		{Ram}(2013)}]{mehta2013fano}%
	\BibitemOpen
	\bibfield  {author} {\bibinfo {author} {\bibfnamefont {K.~K.}\ \bibnamefont
			{Mehta}}, \bibinfo {author} {\bibfnamefont {J.~S.}\ \bibnamefont {Orcutt}}, \
		and\ \bibinfo {author} {\bibfnamefont {R.~J.}\ \bibnamefont {Ram}},\
	}\bibfield  {title} {\enquote {\bibinfo {title} {Fano line shapes in
				transmission spectra of silicon photonic crystal resonators},}\ }\href@noop
	{} {\bibfield  {journal} {\bibinfo  {journal} {Applied Physics Letters}\
		}\textbf {\bibinfo {volume} {102}},\ \bibinfo {pages} {081109} (\bibinfo
		{year} {2013})}\BibitemShut {NoStop}%
	\bibitem [{\citenamefont {Ohta}\ \emph {et~al.}(2011)\citenamefont {Ohta},
		\citenamefont {Ota}, \citenamefont {Nomura}, \citenamefont {Kumagai},
		\citenamefont {Ishida}, \citenamefont {Iwamoto},\ and\ \citenamefont
		{Arakawa}}]{ohta2011strong}%
	\BibitemOpen
	\bibfield  {author} {\bibinfo {author} {\bibfnamefont {R.}~\bibnamefont
			{Ohta}}, \bibinfo {author} {\bibfnamefont {Y.}~\bibnamefont {Ota}}, \bibinfo
		{author} {\bibfnamefont {M.}~\bibnamefont {Nomura}}, \bibinfo {author}
		{\bibfnamefont {N.}~\bibnamefont {Kumagai}}, \bibinfo {author} {\bibfnamefont
			{S.}~\bibnamefont {Ishida}}, \bibinfo {author} {\bibfnamefont
			{S.}~\bibnamefont {Iwamoto}}, \ and\ \bibinfo {author} {\bibfnamefont
			{Y.}~\bibnamefont {Arakawa}},\ }\bibfield  {title} {\enquote {\bibinfo
			{title} {Strong coupling between a photonic crystal nanobeam cavity and a
				single quantum dot},}\ }\href@noop {} {\bibfield  {journal} {\bibinfo
			{journal} {Applied Physics Letters}\ }\textbf {\bibinfo {volume} {98}},\
		\bibinfo {pages} {173104} (\bibinfo {year} {2011})}\BibitemShut {NoStop}%
	\bibitem [{\citenamefont {Liu}\ \emph {et~al.}(2010)\citenamefont {Liu},
		\citenamefont {Pu}, \citenamefont {Yvind},\ and\ \citenamefont
		{Hvam}}]{liu2010high}%
	\BibitemOpen
	\bibfield  {author} {\bibinfo {author} {\bibfnamefont {L.}~\bibnamefont
			{Liu}}, \bibinfo {author} {\bibfnamefont {M.}~\bibnamefont {Pu}}, \bibinfo
		{author} {\bibfnamefont {K.}~\bibnamefont {Yvind}}, \ and\ \bibinfo {author}
		{\bibfnamefont {J.~M.}\ \bibnamefont {Hvam}},\ }\bibfield  {title} {\enquote
		{\bibinfo {title} {High-efficiency, large-bandwidth silicon-on-insulator
				grating coupler based on a fully-etched photonic crystal structure},}\
	}\href@noop {} {\bibfield  {journal} {\bibinfo  {journal} {Applied Physics
				Letters}\ }\textbf {\bibinfo {volume} {96}},\ \bibinfo {pages} {051126}
		(\bibinfo {year} {2010})}\BibitemShut {NoStop}%
	\bibitem [{\citenamefont {Zhang}\ \emph {et~al.}(2018)\citenamefont {Zhang},
		\citenamefont {Leroux}, \citenamefont {Dur\'an-Valdeiglesias}, \citenamefont
		{Alonso-Ramos}, \citenamefont {Marris-Morini}, \citenamefont {Vivien},
		\citenamefont {He},\ and\ \citenamefont {Cassan}}]{zhang2018generating}%
	\BibitemOpen
	\bibfield  {author} {\bibinfo {author} {\bibfnamefont {J.}~\bibnamefont
			{Zhang}}, \bibinfo {author} {\bibfnamefont {X.}~\bibnamefont {Leroux}},
		\bibinfo {author} {\bibfnamefont {E.}~\bibnamefont {Dur\'an-Valdeiglesias}},
		\bibinfo {author} {\bibfnamefont {C.}~\bibnamefont {Alonso-Ramos}}, \bibinfo
		{author} {\bibfnamefont {D.}~\bibnamefont {Marris-Morini}}, \bibinfo {author}
		{\bibfnamefont {L.}~\bibnamefont {Vivien}}, \bibinfo {author} {\bibfnamefont
			{S.}~\bibnamefont {He}}, \ and\ \bibinfo {author} {\bibfnamefont
			{E.}~\bibnamefont {Cassan}},\ }\bibfield  {title} {\enquote {\bibinfo {title}
			{Generating fano resonances in a single-waveguide silicon nanobeam cavity for
				efficient electro-optical modulation},}\ }\href@noop {} {\bibfield  {journal}
		{\bibinfo  {journal} {ACS Photonics}\ }\textbf {\bibinfo {volume} {5}},\
		\bibinfo {pages} {4229--4237} (\bibinfo {year} {2018})}\BibitemShut {NoStop}%
\end{thebibliography}

\end{document}